\documentclass{article}
\usepackage{waspaa21,amsmath,amssymb,graphicx,url,times}
\usepackage{color}
\usepackage{multirow,booktabs}   
\usepackage{cite}
\usepackage{lipsum}
\usepackage{hyperref}

\newcommand\linesubsec[1]{\vspace{0.8mm}\noindent\textbf{#1 --- }}

\ninept

\title{Filtered noise shaping for time domain room impulse \\ response estimation from reverberant speech}

\name{Christian J. Steinmetz$^{1}$\sthanks{Work done during an internship at Facebook Reality Labs Research.} \quad
      Vamsi Krishna Ithapu$^{2}$ \quad
      Paul Calamia$^{2}$}
\address{$^1$Centre for Digital Music, Queen Mary University of London, London, UK\\ 
         $^2$ Facebook Reality Labs Research, Redmond, Washington, USA
        \\       
}

\begin{document}

\ninept
\maketitle

\begin{sloppy}

\begin{abstract}
Deep learning approaches have emerged that aim to transform an audio signal so that it sounds as if it was recorded in the same room as a reference recording, with applications both in audio post-production and augmented reality.
In this work, we propose FiNS, a \textbf{Fi}ltered \textbf{N}oise \textbf{S}haping network that directly estimates the time domain room impulse response (RIR) from reverberant speech.
Our domain-inspired architecture features a time domain encoder and a filtered noise shaping decoder that models the RIR as a summation of decaying filtered noise signals, along with direct sound and early reflection components.
Previous methods for acoustic matching utilize either large models to transform audio to match the target room or predict parameters for algorithmic reverberators.
Instead, blind estimation of the RIR enables efficient and realistic transformation with a single convolution. 
An evaluation demonstrates our model not only synthesizes RIRs that match parameters of the target room, such as the $T_{60}$ and DRR, but also more accurately reproduces perceptual characteristics of the target room, as shown in a listening test when compared to deep learning baselines.
\end{abstract}

\vspace{-0.1cm}
\section{Introduction}
\label{sec:intro}
\vspace{-0.1cm}
The room impulse response (RIR) precisely characterizes the acoustic signature of a space. 
To that end, the RIR drives many applications in acoustical analysis and signal processing, 
such as dereverberation~\cite{makino1993exponentially}, room volume estimation~\cite{kuster2008reliability}, speech recognition~\cite{gomez2008distant}, 
as well as generation of virtual sound sources in augmented and virtual reality applications~\cite{jot2016augmented}. 
Direct measurement of the RIR is often challenging in many environments, 
requiring intrusive test signals, the absence of background noise, and high-fidelity reproduction and capture equipment~\cite{stan2002comparison}. 
These limitations have led to an interest in estimation of acoustic parameters that characterize the room, 
such as the Reverberation Time ($T_{60}$) and Direct-to-Reverberant Ratio ($\text{DRR}$)~\cite{wen2008blind, doire2015single}, 
bypassing the full scale measurement. 
This is often achieved by using unobtrusive excitation signals, such as speech, captured with consumer-grade equipment~\cite{eaton2016estimation}. 

While these parameters provide useful information about room acoustics, they can be limiting in cases where auralization of the RIR is required. 
This is especially the case with augmented and virtual reality (AR and VR), 
and is also the case with reverb matching procedures for audio post-production~\cite{peters2012matching, su2020acoustic, sarroff2020blind, koo2021reverb}.
In these cases, accurately matching the characteristics of the room acoustics is generally required, 
which often extends beyond $T_{60}$ and $\text{DRR}$.
Some additional attributes include room geometry~\cite{kim2020acoustic}, material properties~\cite{schissler2017acoustic}, 
and early reflections~\cite{shlomo2021blind}, all of which are required to simulate and transform a given input signal to a specific target room signature.
Unless accurate estimates of all these attributes are obtained one cannot simulate sounds in general scenes. 

\begin{figure}
    \centering
    \vspace{-0.1cm}
    \includegraphics[width=\linewidth,trim={0.2cm 0 0.3cm 0.1cm},clip]{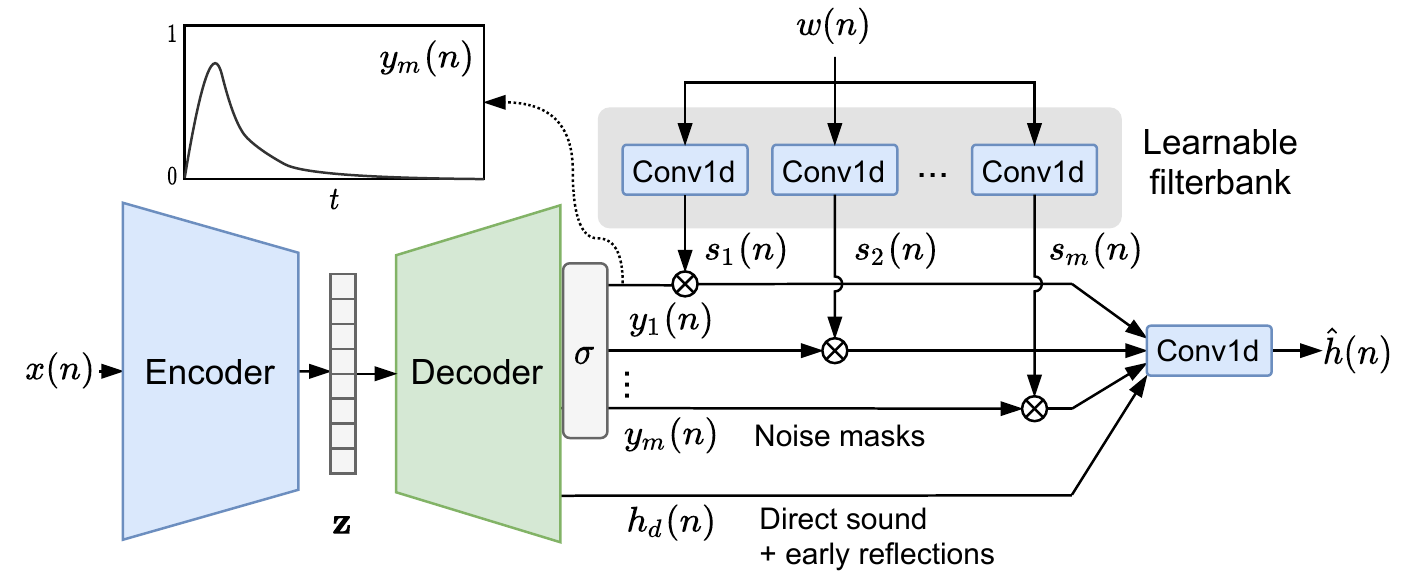}
    \vspace{-0.3cm}
    \caption{\textbf{FiNS}: Filtered noise shaping  RIR synthesis network.}
    \vspace{-0.3cm}
    \label{fig:arch}
\end{figure}

The challenge in estimating these acoustic parameters in conjunction with using them to drive an accurate room simulation 
motivates the design of a system that can directly model the acoustics of the space given only a reference recording.
Several deep learning approaches have been introduced to address this {\it acoustic matching} task.
Many are focused on audio post-production, and therefore rely on training data generated with algorithmic reverbs that may not generalize to real rooms~\cite{peters2012matching, sarroff2020blind, koo2021reverb}. 
Other approaches are trained using measured RIRs and additionally handle removing room effects from the source recordings, but operate at lower sample rates, 
and employ a large model for the transformation, which can be compute intensive during inference~\cite{su2020acoustic}.

In this work, we propose a framework to mitigate these issues. 
We begin by considering the task of modeling the room as a linear time-invariant system, 
where the observed reverberant speech is the result of the convolution between anechoic speech and the RIR. 
Estimation of the RIR based upon the reverberant speech will allow us to, in principle, 
replace room simulations or large transformation models with a simple convolution operation of an input signal and a predicted RIR.
To construct a model for this task we take inspiration from room acoustics, 
where it is common to decompose the RIR into the direct sound, early reflections, and late reverberation. 

We propose \textbf{FiNS}, a \textbf{Fi}ltered \textbf{N}oise \textbf{S}haping network that reconstructs RIRs 
by cascading a time domain encoder and a specialized decoder that models the RIR as a summation of decaying filtered noise signals,
along with components for the direct sound and early reflections.
To aid in applications in post-production and augmented reality, our proposed model produces high-fidelity outputs at 48\,kHz, enabling direct application in AR and VR systems and audio post-production.
Through systematic objective and subjective evaluations, we show that FiNS generates RIRs that match acoustic characteristics of the target room, 
and produces more realistic sounding RIRs, outperforming other deep learning baselines.

\begin{figure}
    \centering
    \vspace{-0.3cm}
    \includegraphics[width=0.94\linewidth,trim={0.1cm 0.05cm 0.6cm 0.6cm},clip]{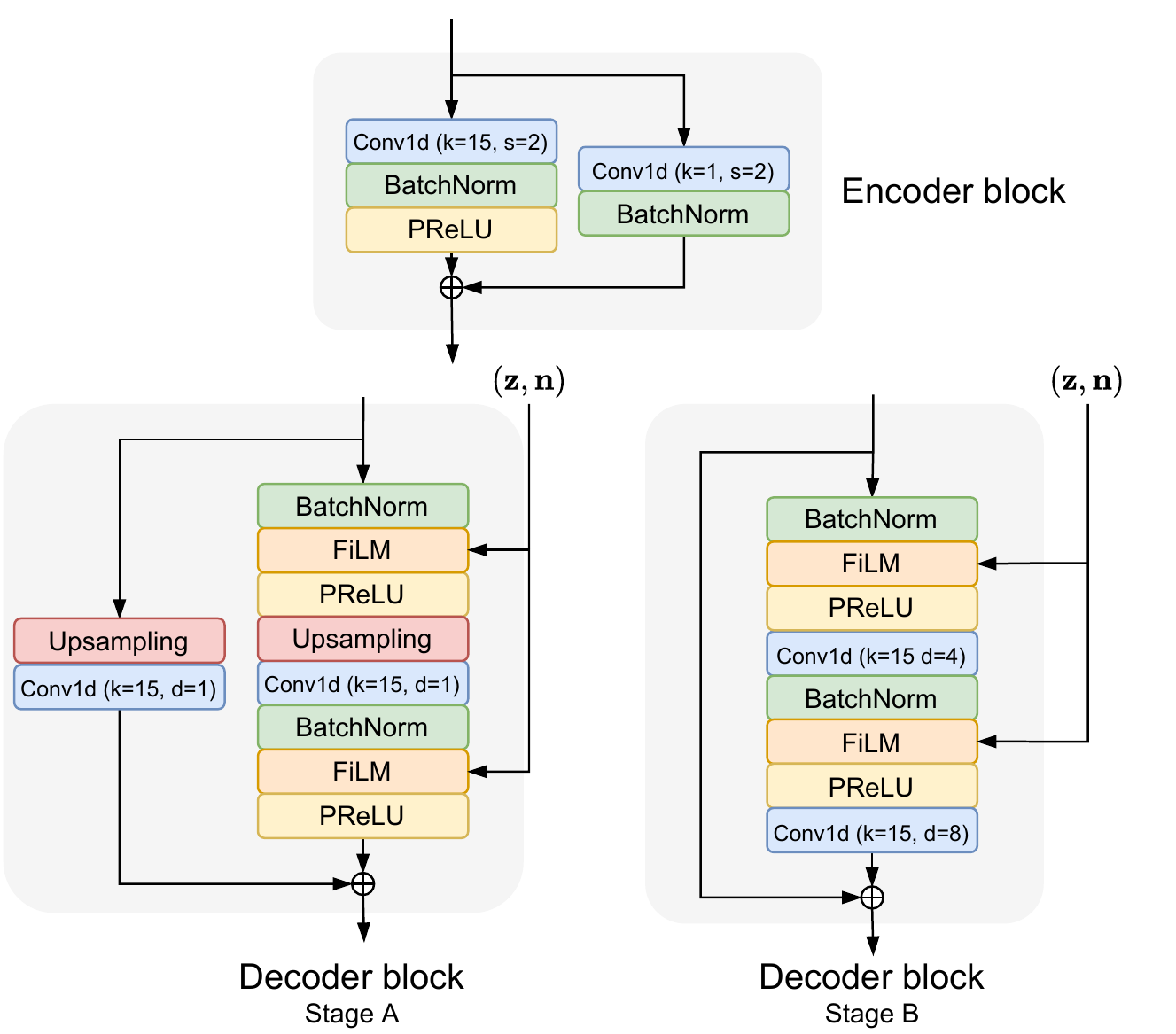}
    \vspace{-0.2cm}
    \caption{Encoder and decoder block structures.}\label{fig:blocks}
    \vspace{-0.4cm}
\end{figure}

\section{Related Work}

Over the past decade, a number of signal processing approaches for the blind estimation of acoustic room parameters 
have been proposed~\cite{wen2008blind, falk2010temporal, doire2015single}.
More recently, deep learning approaches have brought competitive performance, but are often challenged by limited training data, 
requiring underparameterized models~\cite{gamper2018blind, xiong2018joint}, or data augmentation~\cite{bryan2020impulse} to avoid overfitting.
While these methods continue to improve, the success of end-to-end deep learning models has led to an interest in the direct modeling of room acoustics. 

For the task of acoustic matching for audio post-production, Sarroff and Michaels~\cite{sarroff2020blind} proposed an RNN trained to predict the parameters of an algorithmic reverb given a reverberant recording. 
Koo \emph{et al.}~\cite{koo2021reverb} extended this with a U-Net trained to directly convert singing voice recordings to match a target signal, forgoing the need to estimate the parameters of an algorithmic reverb.
Nevertheless, while these results are convincing, both methods rely upon training data generated with algorithmic reverbs, which may limit generalization for matching real environments in augmented reality.

The approach proposed by Su \emph{et al}.~\cite{su2020acoustic} is the most similar to ours. 
Their architecture is composed of a spectral encoder and conditional feedforward WaveNet, 
which processes a signal in the time domain, transforming it to the target room based upon a room embedding.
While they demonstrate the ability to match the characteristics of real rooms and additionally handle the removal of reverberation in the source signal, 
their method operates at 16\,kHz, and  due to the large 
feedforward WaveNet network used in the transformation process.

In addition to the acoustic matching task, 
there has also been interest in the generation of RIRs with applications in data augmentation and training of speech processing models.
Ratnarajah \emph{et al.}~\cite{ratnarajah2020ir} proposed the use of GANs for synthesizing RIRs, 
using interpolation in the latent space of the model to control attributes of synthesized RIRs.
Wager \emph{et al.}~\cite{wager2020dereverberation} also utilized the prediction of the RIR magnitude 
response as a further loss term while training a dereverberation model. 
There has also been significant work in the construction of realistic artificial reverberators~\cite{valimaki2012fifty}, 
with techniques like feedback delay networks~\cite{rocchesso1997circulant} and velvet noise~\cite{valimaki2017late}.
While these techniques offer efficient methods for generating plausible artificial reverberation, 
blind automatic parameterization of these methods to match a target room remains an open research question. 


\section{Approach}

We denote the anechoic speech, the RIR, and the corresponding reverberant speech signal as $x(n), h(n)$ and $x_r(n)$ respectively. 
Here $x(n), x_r(n) \in \mathbb{R}^{1 \times T}$ and $h(n) \in \mathbb{R}^{1 \times L}$ 
($T$ and $L$ denote the number of samples in the speech and RIR, where $n$ is the sample index).
\subsection{Encoder}

The input $x_r(n)$ is processed by a time domain encoder composed of strided 1-D convolutions, as depicted in Figure~\ref{fig:arch},
progressively downsampling the signal to produce a $D$-dimensional latent embedding $\textbf{z} \in \mathbb{R}^{1 \times D}$. 
A time domain encoder is motivated by the need to capture behavior at both small and large time scales within the RIR, which otherwise may require more complicated multi-scale spectral features.  
The composition of each convolutional block is shown in the top of Figure~\ref{fig:blocks}, 
consisting of a 1-D convolution with a kernel size of 15 and a stride 2, followed by batch normalization, 
and a parameterized ReLU (PReLU) activation~\cite{he2015delving}.
Residual connections are implemented with $1 \times 1$ convolutions using the same stride, followed by batch normalization. 
As is common, the number of channels in each convolutional block is progressively increased through the depth of the network, 
producing an output with 512 channels at the final layer.
At the output of the convolutional blocks, adaptive average pooling is utilized to aggregate information across time 
for utterances larger than the receptive field.
This produces a fixed size embedding of 512 dimensions, which is passed to a three-layer MLP to produce $\textbf{z}$, of 128 dimensions.
With a total of fourteen blocks, the encoder achieves a receptive field of over 100,000 timesteps, which equates to 2.4 seconds at 48\,kHz.


\subsection{Decoder}

We construct a decoder with strong inductive bias for RIR synthesis in an effort to produce more realistic sounding outputs given limited training data. 
We first consider that it is common to decompose the RIR into three components: 
the direct sound, the early reflections, and the late reverberation~\cite{valimaki2012fifty}.
Interestingly, the late reverberation can be modeled with exponentially decaying filtered noise,
first noted by Moorer over 40 years ago~\cite{moorer1979reverberation}.
Inspired by this, instead of constructing a decoder that directly synthesizes each sample of the entire RIR, 
we model the late reverberation as a sum of filtered noise signals, shaped by time domain masks produced by the decoder.
We achieve this by training the decoder to generate $M$ time domain masks $y_1(n), ...,y_m(n) \in \mathbb{R}^{1 \times L}$ 
that are applied to a set of filtered noise signals $s_1(n), ...,s_m(n) \in \mathbb{R}^{1 \times L}$, 
such that each subband of the late signal $\hat{h}_{l,m}(n)$ is the element-wise product
\begin{equation*}\label{eq:late}
    \hat{h}_{l,m}(n) = \sigma \big( y_m \left(n\right)  \big) \odot s_m(n),
\end{equation*}
where $\sigma(\cdot)$ is the sigmoid function and $L$ is the number of samples in the RIR.
To enable control over the spectral content, the filtered noise signals $s_{m}(n)$ are generated by passing a noise signal $w(n)$ through a trainable filterbank containing $M$ FIR filters of order $N$, implemented with 1-D convolutions.
Each filtered noise signal $s_{m}(n)$ is therefore given by the following convolution, 
\begin{equation*}
    s_{m}(n) = \sum_{i=0}^{N} b_{m,i} \cdot w[n-i],
\end{equation*}
where $b_{m,i}$ is the $i$-th coefficient of the $m$-th filter.
This structure is demonstrated on the left of Figure~\ref{fig:arch}, where the masks from the decoder are applied to a series of filtered noise signals.

While this enables the ability to model the late part of the reverberation, 
the direct sound and early reflections cannot accurately be captured in this manner. 
To account for these, we simply allocate an extra output channel of the decoder $h_d(n) \in \mathbb{R}^{1 \times E}$ 
to directly predict the time domain response, where $E$ corresponds to the number of samples in the early component.
This is shown as the bottommost output of the decoder in Figure~\ref{fig:arch}.
In practice, we zero all samples in $h_d(n)$ where $n > E$, with $E=2400$, corresponding to the first 50\,ms.
The RIR, $\hat{h}(n)$, is produced by mixing the $M$ late components $h_{l,m}(n)$ and the early component $h_d(n)$ with a $1 \times 1$ convolution to produce the final monophonic RIR.

The decoder architecture is based on the generator in \mbox{GAN-TTS}~\cite{binkowski2019gantts}, a feedforward speech synthesis model.
As shown in the bottom of Figure~\ref{fig:blocks}, each block is comprised of two stages. 
The first stage features upsampling with transposed convolutions, 
and the second stage further refines the upsampled signal with additional convolutions using larger dilation factors.
FiLM~\cite{perez2018film} conditioning is used in both stages to continually inject information from the latent embedding $\textbf{z}$, which is concatenated with a noise vector $\textbf{n} \in \mathbb{R}^{1 \times D}$.
Linear layers in each FiLM operation project this conditioning vector to the proper number of gain and shift parameters. 
While the input to GAN-TTS is a vector of linguistic features, 
we instead utilize a single trainable vector $\textbf{v} \in \mathbb{R}^{1 \times K}$ where $K=400$.

We found proper initialization of the filterbank to be critical in achieving convergence. 
We experimented with both dirac and octave-spaced bandpass filter initializations and found that the bandpass initialization appeared to work best 
(standard Kaiming~\emph{et~al.}~\cite{he2015delving} initialization resulted in poor convergence).
In our final model, we utilized $M=10$ filtered noise signals, with filters of order $P=1023$ to provide sufficient low frequency response.

\subsection{Loss functions}

Selecting a loss function for training is challenging due to the differing structure within the components of the RIR. 
The early components tend to exhibit sparse, impulsive character, while the late components are noise-like.
We experimented with a number of formulations including both time domain and spectral distances, and their combination, 
yet we found the multiresolution STFT loss~\cite{yamamoto2020parallel} alone produced the best results.
This loss is composed of two terms, the spectral convergence $\mathcal{L}_{\text{SC}}$ (\ref{eq:spectral-convergence}), 
and spectral log-magnitude $\mathcal{L}_{\text{SM}}$ (\ref{eq:spectral-magnitude}), where $||\cdot||_{\text{F}}$ is the Frobenius norm, 
$||\cdot||_1$ is the $L_1$ norm, and $N$ is the number of STFT frames. 
The multiresolution STFT loss $\mathcal{L}_{\text{STFT}}$ is then computed as the sum of these two terms, over $R$ different STFT resolutions, as shown in (\ref{eq:stft-loss}).
\begin{align} 
    \mathcal{L}_{\text{SC}}({\hat{h}},h) &= \frac{\| ~|\text{STFT}(h)| - |\text{STFT}(\hat{h})|~ \|_{\text{F}}}{\| ~|\text{STFT}(h)|~ \|_{\text{F}}} \label{eq:spectral-convergence} \\ 
    \mathcal{L}_{\text{SM}}({\hat{h}},h) &= \frac{1}{N} \left\| \log \left( ~|\text{STFT}(h)|~ \right) - \log ( ~|\text{STFT}(\hat{h})|~ ) \right\|_1 \label{eq:spectral-magnitude} \\
    \mathcal{L}_{\text{STFT}}({\hat{h}},h) &= \sum_{r=1}^{R} \mathcal{L}_{\text{SC}_{r}}({\hat{h}},h) + \mathcal{L}_{\text{SM}_{r}}({\hat{h}},h) \label{eq:stft-loss}
\end{align}




\section{Evaluation}
\label{sec:eval}

\linesubsec{Baselines}
We construct a simple deep-learning baseline by adapting \mbox{Wave-U-Net}~\cite{stoller2018waveunet}, 
a popular time domain autoencoder network architecture originally designed for source separation. 
We largely follow the original architecture, using a total of 12 layers in the encoder and decoder, 
growing the number of convolutional channels in each layer by 24. 
We also replace the linear upsampling in the decoder with transposed convolutions. 
Unlike FiNS, \mbox{Wave-U-Net} includes skip connections between the encoder and decoder. 
While these have been shown to aid convergence, 
we posit that they are likely not helpful due to the disconnection between the time scales of the input speech utterance and the output RIR.
Since Wave-U-Net is deterministic, it may struggle to accurately reproduce the noise-like late reverberation.
Further, in order to evaluate the efficacy of the noise shaping decoder, 
we also compare with a variant of our proposed model where the noise shaping component is removed, called \mbox{FiNS (D)}.
We then train the decoder to directly predict every sample of the RIR, as is common in audio synthesis architectures~\cite{binkowski2019gantts}. 
While the noise shaping masks are not employed in \mbox{FiNS (D)}, 
noise vectors $\textbf{n}$ are still utilized within each FiLM operation, and all other model hyperparameters are kept the same. 

\linesubsec{Training}
We utilize VCTK~\cite{yamagishi2019vctk} as a source of clean speech,
which provides around 80\,k studio recorded utterances from 110 English speakers at 48\,kHz,
and an FRL dataset of measured RIRs. 
We create a training (80\%), validation (10\%), and test split (10\%) across these datasets avoiding overlapping speakers and rooms. 
During training, we generate reverberant signals by sampling utterances and RIRs from the training set, 
producing inputs of 131072 samples ($\approx$2.73\,s at 48\,kHz).
Target RIRs are made 48,000 samples (1 second) in length by padding or cropping where appropriate, 
with all models operating on input and output audio at 48\,kHz.
Each model is trained for a total of 300 epochs with a batch size of 128 using AdamW~\cite{loshchilov2017decoupled}. 
We set an initial learning rate of $0.0001$, decreasing it by half every 40k steps. 
We apply gradient clipping, restricting the norm of the gradients to 5.
Additionally, we used automatic mixed precision. 
For the multiresolution STFT loss, we utilize $R=4$ resolutions with both small and large frame sizes of 64 (1.3\,ms), 512 (10.6\,ms), 2048 (42.7\,ms), and 8192 (170.6\,ms), covering a range of time scales present within common RIRs. 
The hop size for each resolution is always half the frame size. 

\begin{figure}[t]
\minipage{0.23\textwidth}
    \centering
  \includegraphics[width=\linewidth]{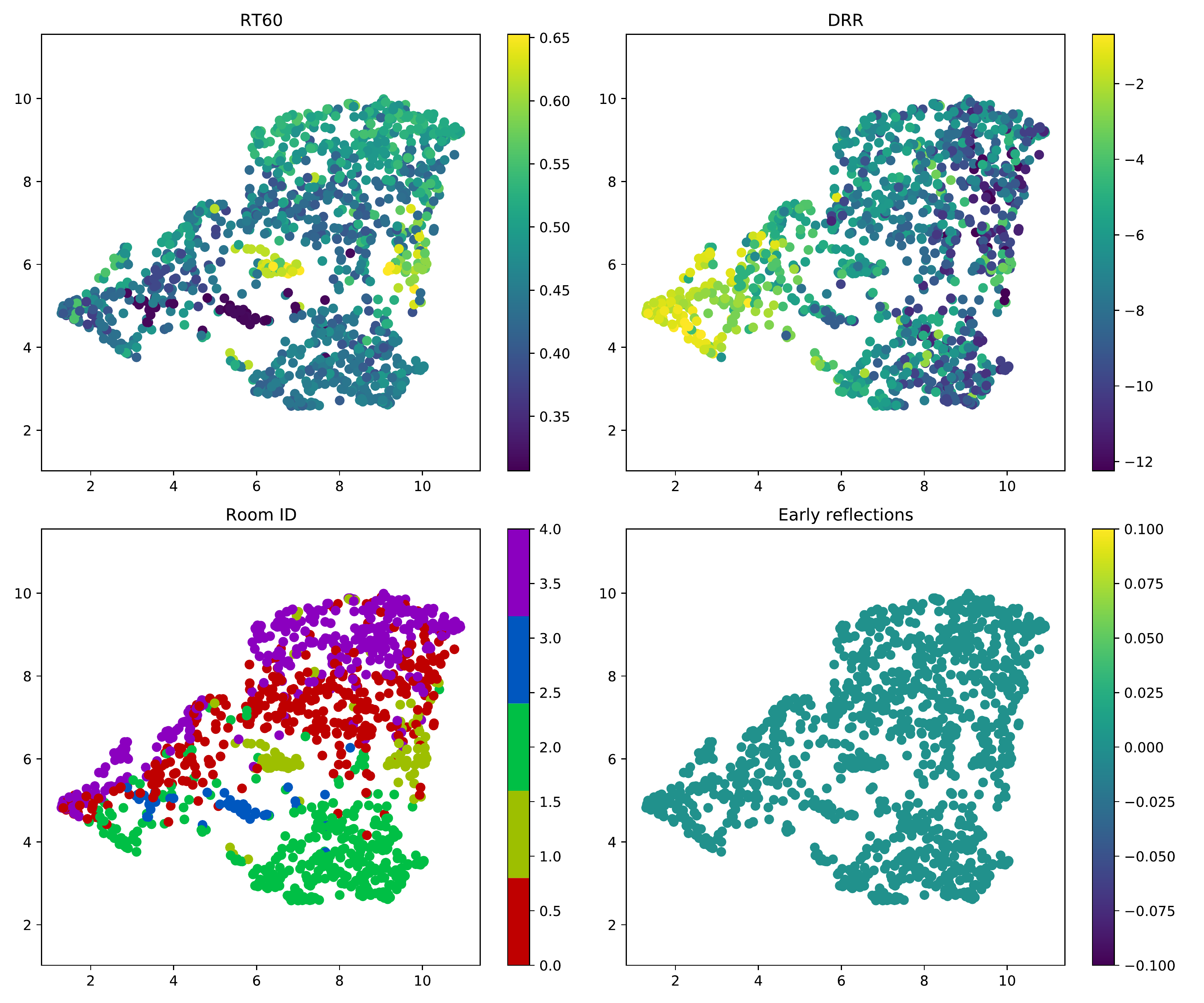}
     \textbf{(a)}
    \vspace{-0.2cm}
\endminipage\hfill
\minipage{0.23\textwidth}
    \centering
  \includegraphics[width=\linewidth]{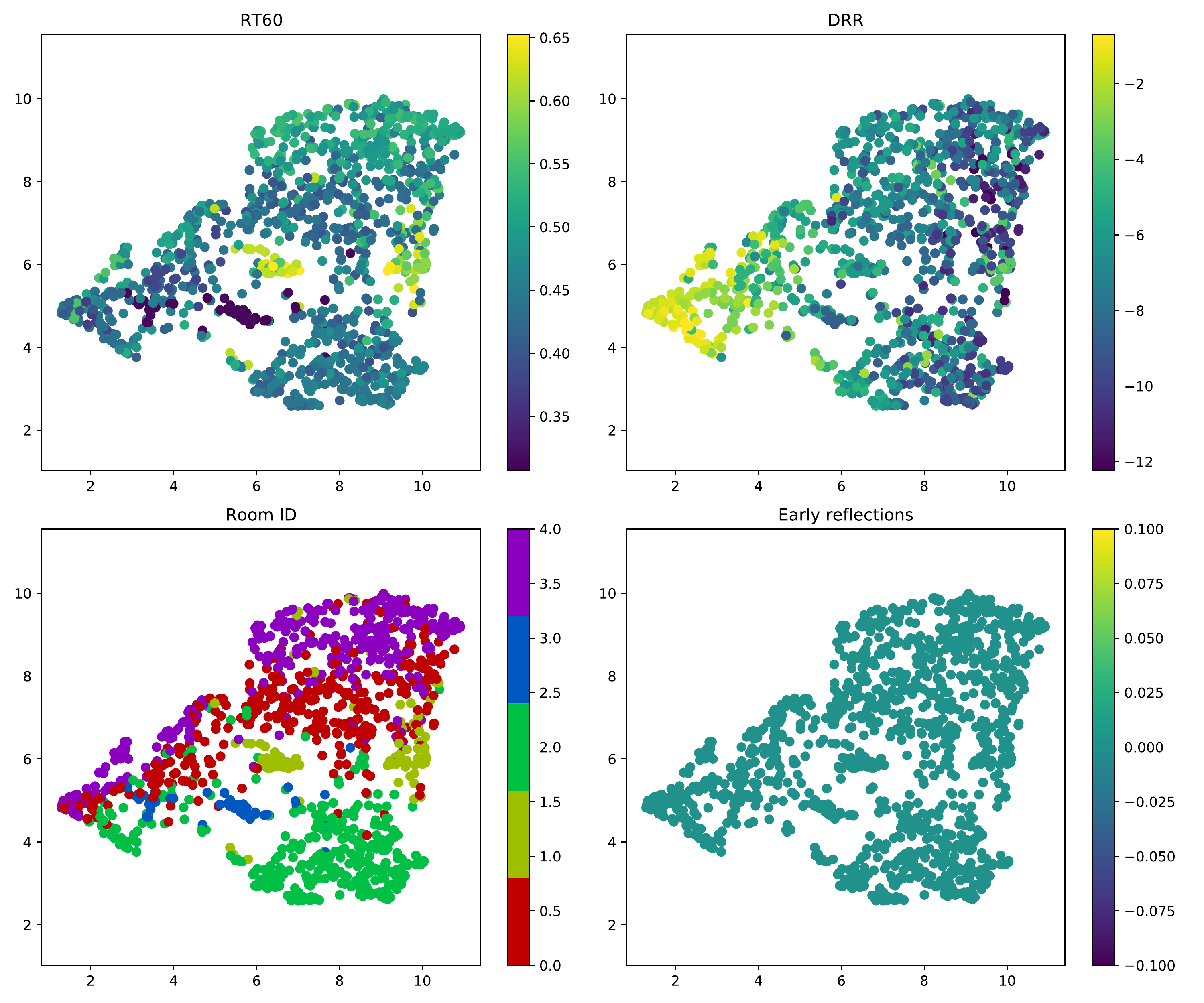}
     \textbf{(b)}
  \vspace{-0.2cm}
\endminipage
 \caption{2-D projections of embeddings from the encoder for unseen examples, colored by (a) ground-truth DRR and (b) room ID.}\label{fig:latents}
\vspace{-0.5cm}
\end{figure}

\renewcommand{\arraystretch}{0.8}
\begin{table*}[ht]
  \vspace{-0.3cm}
  \centering
    \begin{tabular} {l l l r rrr rrr}
    \toprule
    \multirow{2}{*}{\textbf{RIR}} & \multirow{2}{*}{\textbf{Speech}} & \multirow{2}{*}{\textbf{Model}}  &
    \multirow{2}{*}{$\mathcal{L}_{\text{STFT}} \downarrow$} & \multicolumn{3}{c}{\boldmath${T_{60}}$} & \multicolumn{3}{c}{\textbf{DRR}}  
    \\  \cmidrule(lr){5-7}  \cmidrule(lr){8-10} 
                                       & &   &  &  \footnotesize{Bias $|\downarrow|$} & \footnotesize{MSE (s) $\downarrow$} & \footnotesize{$\rho \uparrow$}  &  \footnotesize{Bias $|\downarrow|$} & \footnotesize{MSE (dB) $\downarrow$} & \footnotesize{$\rho \uparrow$} 
                                  \\  \midrule
    \multirow{3}{*}{FRL} & \multirow{3}{*}{VCTK} 
     & Wave-U-Net  &  1.127 &  -0.016 & 0.005 & 0.480 &  -0.25 &    4.19 &  0.736 \\
    & & FiNS (D)  &  \textbf{1.064} &  -0.001 & 0.004 & 0.548 &   0.54 &   4.13 &  0.734 \\
    & & FiNS   &  1.157 &   0.041 & 0.005 & 0.646 &   0.43 &    4.45 &  0.721 \\  \midrule
    \multirow{3}{*}{FRL} & \multirow{3}{*}{ACE}
     & Wave-U-Net  &  \textbf{1.119} &   0.006 & 0.004 & 0.495 &  -0.58 &    5.55 &  0.625 \\
    & & FiNS (D)   &  1.137 &   0.034 & 0.006 & 0.479 &   0.50 &   5.14 & 0.661 \\
    & & FiNS   &  1.183 &   0.057 & 0.008 & 0.540 &   0.50 &    6.29 &  0.625 \\  \midrule
        \multirow{3}{*}{ACE} & \multirow{3}{*}{VCTK} 
     & Wave-U-Net  &  2.295 &   0.022 & 0.135 & 0.324 &  -6.32 &    49.28 &  0.681 \\
    & & FiNS (D)  &  \textbf{2.040} &  -0.148 & 0.093 & 0.658 &  -5.67 &   40.33 &  0.750 \\
    & & FiNS   &  2.060 &  -0.068 & 0.056 & 0.837 &  -5.76 &   43.62 &  0.640 \\  
                            \bottomrule
    \end{tabular} 
    \vspace{-0.1cm}
    \caption{Test performance on objective metrics across different models on FRL and ACE Challenge sets.}
    \label{tab:evaluation}   
    \vspace{-0.3cm}
\end{table*}

\subsection{Latent space}

We begin by inspecting embeddings produced by the encoder on combinations of unseen utterances and RIRs. 
We create 2-D projections of the 128-dimensional embeddings using UMAP~\cite{mcinnes2018umap} and overlay both the ground truth DRR and room identity of the RIRs as shown in Figure \ref{fig:latents}.
These plots demonstrate that clear structure exists within the latent space, indicating that the encoder has learned implicitly to capture information about the target room based solely on the reverberant speech.
This is expected, as this information is likely required at a minimum to estimate the RIR. 

\subsection{Objective metrics}

In Table~\ref{tab:evaluation} we report the reconstruction error using the multi-resolution STFT loss, 
$\mathcal{L}_{\text{STFT}}$, as well as the degree to which generated RIRs match the $T_{60}$ and DRR of the target RIR, 
with the bias, MSE, and Pearson correlation coefficient, $\rho$.
This provides a rough indication of the ability to match the characteristics of the target room.
In the first row, using FRL RIRs and speech from VCTK, we find that all models perform similarly and appear to match the room acoustics parameters.
While these metrics appear to suggest all models perform somewhat similarly, 
informal listening indicated that there is often a significant difference in quality among the predicted RIRs. 
Both Wave-U-Net and FiNS (D) produce noticeable unnatural ringing artifacts in the generated RIRs, which are not present in the outputs from FiNS with the noise shaping decoder and not captured by these objective metrics.
We further investigate these perceptual differences in a subjective listening test described in the next section with listening examples made available online\footnote{\url{https://facebookresearch.github.io/FiNS}}.

As a test of generalization, we also report performance on RIRs and speech from the ACE Challenge~\cite{eaton2016estimation}, shown in the second two rows of Table~\ref{tab:evaluation}.
Performance appears similar to before, using RIRs from the FRL dataset and speech from ACE, indicating good generalization for out-of-distribution utterances. 
In the final row, we observe a decrease in performance using RIRs from ACE and speech from VCTK, but we observe that Wave-U-Net is outperformed by both FiNS variants.
Nevertheless, it is clear that the overall performance has decreased predicting RIRs from ACE, likely due to the fact that this dataset contains out-of-distribution RIRs with a wider range of $T_{60}$ and DRR compared to the FRL dataset.

\subsection{Listening test}

Our informal listening tests indicated models achieving similar reconstruction error often sounded significantly different due to the presence of artifacts. 
For this reason, we carried out a more extensive listening test based on the MUSHRA~\cite{mushra} design.
In this test, 15 listeners were tasked with rating the similarity to the reference of speech utterances generated by convolving the predicted RIR from each method with the original anechoic utterance.
Reverberant utterances from the three models, Wave-U-Net, FiNS (D), and FiNS, were included, using ground truth RIRs sourced from the FRL test set, and speech from the VCTK and ACE test sets.
Additionally, a hidden anchor is included, which is an RIR generated by computing the mean RIR across the training set, along with the hidden reference.
Six different utterances were evaluated, with ratings provided for each of the five stimuli, as shown in Figure~\ref{fig:mushra}, with aggregated ratings for each method across all utterances shown on the right.

\begin{figure}
    \centering
    \includegraphics[width=\linewidth,trim={0.1cm 0.0cm 0.3cm 0.0cm},clip]{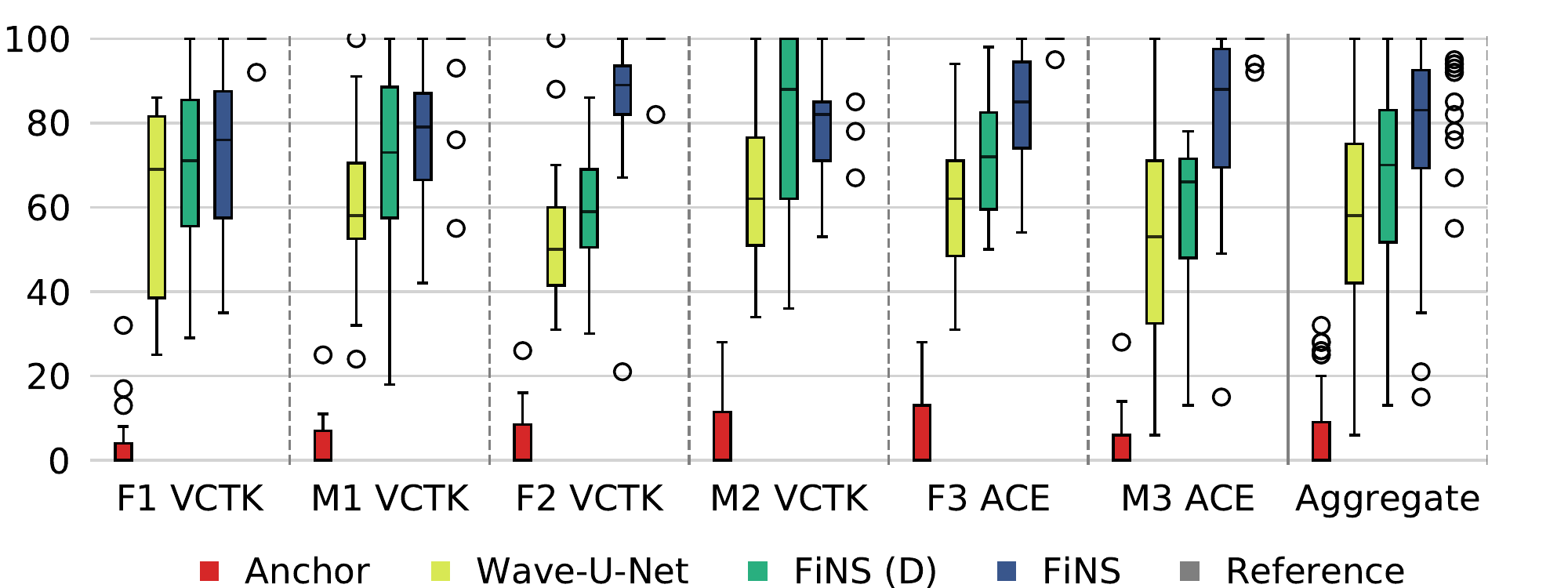}    
    \vspace{-0.5cm}
    \caption{Listening test results.}
    \label{fig:mushra}
    \vspace{-0.5cm}
\end{figure}

Visual inspection indicates a trend with FiNS outperforming the other methods in five out of the six cases.
While listeners generally identified the hidden reference, often rating it the highest, the large range of ratings for the hidden reference in some trials indicates there are cases where the differences are likely subtle.
To formalize these findings, we perform the Kruskal–Wallis $H$ test on the aggregated ratings across all utterances.
This indicates a significant difference among the medians of the groups ($F = 653.9, p \ll 0.001$).
We perform a post-hoc analysis with Conover's test with the Holm adjustment, which reveals a significant difference among all pairwise comparisons of the 
groups, with ($p_{\text{adj}} \ll 0.001$) between FiNS and the hidden reference and ($p_{\text{adj}} \ll 0.001$) between FiNS and FiNS (D).
These results indicate that listeners rated RIRs produced by FiNS the most similar to the reference, yet they could still differentiate among them.

\section{Conclusion}
\label{sec:con}

We proposed FiNS, a filtered noise shaping model for estimation of the time domain room impulse response given a reverberant speech utterance.
The model features a convolutional time domain encoder with an RIR synthesis decoder inspired by knowledge of room acoustics.
Instead of directly predicting each sample of the RIR, the decoder models the synthesis of RIRs as a combination of decaying filtered noise signals, including components for the direct sound and early reflections. 
Objective metrics indicated that the proposed model can accurately capture information about the target room based upon a speech utterance.
A listening test further demonstrated that our approach produces results more perceptually similar to the target room as compared to a Wave-U-Net baseline and a variant of our model without the noise shaping decoder.
While these results are promising, they also suggest estimating the RIR of rooms with $T_{60}$ and DRR outside the training distribution remains challenging.
Future directions include the integration of data augmentation, specialized adversarial losses, as well as multimodal approaches combining audio/visual input from the target room~\cite{singh2021image2reverb}.


\clearpage
\bibliographystyle{IEEEtran}
\bibliography{references}

\end{sloppy}
\end{document}